\documentclass[conference]{IEEEtran}
\IEEEoverridecommandlockouts
\usepackage[pdftex]{graphicx}
\usepackage[ruled]{algorithm2e}
\usepackage{algorithmic}
\usepackage[small]{caption}
\usepackage{float}
\usepackage{xspace}
\usepackage{fancyhdr}
\usepackage{amssymb,amsmath}
\usepackage{subfig}
\usepackage{flushend}
\usepackage{textcomp} 
\usepackage{epstopdf}
\usepackage{bbding}
\usepackage{cite} 
\usepackage{color}
\usepackage{pifont}
\begin{document}

\title{{\huge When RIS Meets GEO Satellite Communications: A \\New Sustainable Optimization Framework in 6G}}
\author{Wali Ullah Khan, Eva Lagunas, Asad Mahmood, Basem M. ElHalawany,\\ Symeon Chatzinotas, Bj\"orn Ottersten \\Interdisciplinary Centre for Security, Reliability and Trust (SnT), University of Luxembourg\\ \{waliullah.khan, eva.lagunas, asad.mahmood, symeon.chatzinotas, bjorn.ottersten\}@uni.lu,\\ basem.mamdoh@feng.bu.edu.eg 

}%
\markboth{}%
{Shell \MakeLowercase{\textit{et al.}}: Bare Demo of IEEEtran.cls for IEEE Journals} 

\maketitle

\begin{abstract}
Reflecting intelligent surfaces (RIS) is a low-cost and energy-efficient solution to achieve high spectral efficiency in sixth-generation (6G) networks. The basic idea of RIS is to smartly reconfigure the signal propagation by using passive reflecting elements. On the other side, the demand of high throughput geostationary (GEO) satellite communications (SatCom) is rapidly growing to deliver broadband services in inaccessible/insufficient covered areas of terrestrial networks. This paper proposes a GEO SatCom network, where a satellite transmits the signal to a ground mobile terminal using multicarrier communications. To enhance the effective gain, the signal delivery from satellite to the ground mobile terminal is also assisted by RIS which smartly shift the phase of the signal towards ground terminal. We consider that RIS is mounted on a high building and equipped with multiple re-configurable passive elements along with smart controller. We jointly optimize the power allocation and phase shift design to maximize the channel capacity of the system. The joint optimization problem is formulated as nonconvex due to coupled variables which is hard to solve through traditional convex optimization methods. Thus, we propose a new $\epsilon-$optimal algorithm which is based on Mesh Adaptive Direct Search to obtain an efficient solution. Simulation results unveil the benefits of RIS-assisted SatCom in terms of system channel capacity.
\end{abstract}


\IEEEpeerreviewmaketitle

\section{Introduction} 
Due to the rapid growth of Internet users and broadband services, communication technologies need to be designed in such a way that cope with the exponential growth of heterogeneous devices and their diverse service demands in the sixth-generation (6G) era \cite{9397776,9479745}. According to a recent study, more than 60\% of the world population are using Internet services which have taken the total number to 4.72 billion by April 2021 \cite{DigGlob}. The study has also reported over the 7\% increase in the last 12 months, as a result, 332 million people came online through the Internet for the first time. However, restrictions related to COVID-19 continue to hamper research into internet adoption around the world. As a result, it is possible that actual growth is been even faster because internet access has become even more important during the pandemic. 

The recent development in modern aerospace technologies able satellite communications (SatCom) to provide heterogeneous services in various application scenarios such as broadcasting, navigation, and disaster relief \cite{9450017}. As a result, the number of connected ground terminals is increasing rapidly. However, the increasing demand for the already limited spectrum resources and energy reservoirs are some of the main challenges \cite{9460776}. Besides that, the effect of obstacles and shadowing between satellite and ground users can disrupt the line-of-sight communication, which may lead to a significant decrease in system performance. To this end, there are demands of high spectral and energy-efficient technologies to extend the wireless coverage and support the explosive growth of ground terminals and their diverse quality of services. In this regard, there emerge some advance technologies such as non-orthogonal multiple access \cite{9468352}, reflecting intelligent surfaces (RIS) \cite{9424177}, and backscatter communications \cite{9328505}.

As considered one of the promising technologies for 6G, RIS has attracted a lot of interest from both academia and industry due to high spectral and energy efficiency \cite{9353406}. The basic idea of RIS is to reflect the signals with adjustable phase shifts towards specific points where needed. These surfaces are very thin which can be coated over building walls, ceilings, and other places. It can be tuned electronically through a software controller using electromagnetic properties. The main features of RIS are the extension of wireless coverage, ensuring user fairness, improving data security, cost-effective, and low hardware complexity, and implementation cost. Inspired by these advantages, it is valuable to integrate SatCom with RIS to further improve the system throughput, coverage, and connectivity. At present, most of the research works on RIS-assisted networks focus on various terrestrial scenarios such as multi-input-multi-output \cite{9127834}, millimeter-wave communications \cite{9110835}, unmanned aerial vehicle \cite{9209992}, device-to-device communications \cite{9301375}, non-orthogonal multiple access \cite{9345507}, backscatter communications \cite{9448441}, cognitive radio \cite{9235486}, cooperative communications \cite{9003222}, and intelligent transportation systems \cite{9144463}. 

Recently, some researchers have also integrated RIS in SatCom networks. For example, Xu {\em et al.} \cite{9343764} have considered satellite-terrestrial cognitive radio network. In the primary network, a satellite communicates with the primary ground user in the presence of an eavesdropper. A secondary transmitter communicates with the secondary receiver through a direct link and RIS-assisted link in the secondary network. Meanwhile, the secondary transmitter sends jamming signals to the eavesdropper by direct link and RIS-assisted link to minimize its received signal strength. Thus, their objective is to minimize the signal to interference plus noise ratio (SINR) at eavesdropper while ensuring the minimum SINR of primary and secondary users. The joint optimization problem is formulated as nonconvex first and then transformed into two subproblems. An alternation optimization algorithm for joint beamforming and reflection phase shift is designed to obtain an efficient solution. Another study in \cite{9200674} has considered low earth orbit (LEO) SatCom network assisted by RIS. They develop a continuous-time model for the multipath channel of moving receiver and investigate the configuration of RIS under delay spread, Doppler spread and received power. The authors have shown that phase shift at RIS can be selected such that the received power is maximized without incurring the Doppler spread to the system. Of late, the authors of \cite{tekbiyik2020reconfigurable} have considered LEO SatCom, where inter-satellite links are assisted by RIS using the terahertz band. They first investigate the error performance in inter-satellite links due to the misalignment fading. Then, the authors use RIS mounted on the neighboring satellite to compensate for the high path loss and improve the signal propagation. Later, they also derive the closed-form of error rate expressions to check the impact of RIS on the misalignment fading. 
\begin{figure}[H]
\centering
\includegraphics [width=0.46\textwidth]{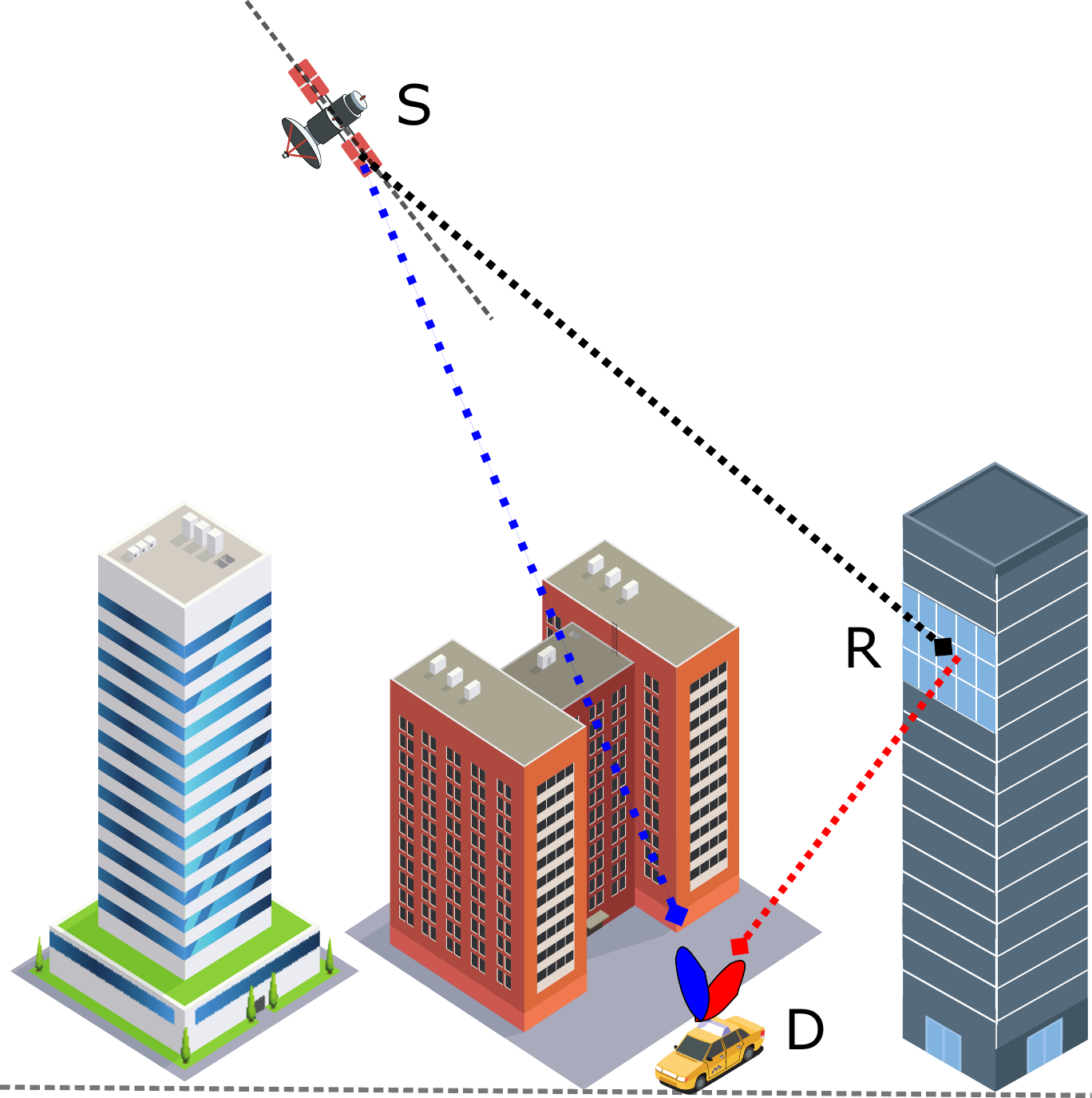}
\caption{System model of RIS-assisted satellite communications.}
\label{Sat}
\end{figure}

Although the above works in \cite{9343764,9200674,tekbiyik2020reconfigurable} have investigated various aspects of the system, they only focus on non-geostationary orbit (GEO) SatCom networks and the work that considers RIS-assisted GEO SatCom network has not yet been studied, to the best of our knowledge. To fill this gap, we investigate the performance of RIS-assisted GEO SatCom network, where the signal from satellite terminal can be received through both direct and assisted links. More specifically, we provide a joint optimization framework to maximize the channel capacity of the system. Our framework simultaneously optimizes the transmit power of the satellite and the reflecting phase shift of RIS while ensuring the minimum channel capacity. The main contributions of this paper can be summarized as follows:
\begin{itemize}
    \item We consider a Geo SatCom which transmits data to the ground terminal through multicarriers. To enhance the signal gain, we consider direct link and RIS-assisted link, where RIS is mounted on the top of the building and assist the signal delivery from satellite to the ground terminal. 
    \item We formulate a joint optimization problem of power allocation and reflecting phase shift design to maximize the channel capacity of the system. Since the problem is nonconvex and hard to solve through conventional convex optimization methods, we apply Mesh Adaptive Direct Search (MADS) algorithm to obtain an efficient solution.
    \item We also provide the numerical results based on Monte Carlo simulations. The proposed RIS-assisted SatCom network is compared with the conventional SatCom without RIS assistance. The plotted results unveil the benefits of RIS in SatCom in the form of high channel capacity compared to its counterpart conventional network.
\end{itemize}
\begin{table}[h]
\centering
\caption{Different symbols and their definitions}
\begin{tabular}{|c||c|} 
\hline 
Symbol & Definition  \\
\hline
${\textit{\textbf{g}}_{m,k}}$ & Channels from satellite to RIS\\\hline 
${\textit{\textbf{h}}_{m,k}}$ & Channels from RIS to ground mobile terminal\\\hline 
$f_m$ & Channel from satellite to ground mobile terminal\\\hline 
$p_t$ & Total transmit power of satellite terminal\\\hline 
$p_m$ & Transmit power over $m$-th subcarrier\\\hline 
${\bf \Theta}$ & Reflection coefficients of RIS\\\hline 
${G^S_m}$ & Beam gain over $m$-th subcarrier\\\hline 
$G^D_m$ & Antenna gain of ground mobile terminal\\\hline 
$X_R,Y_R$ & Width and length of RIS\\\hline 
$d_1$ & Width and length of RIS\\\hline  
$\psi$ & Angle of incidence at RIS\\\hline 
$\varphi$ & Antenna phase at satellite terminal\\\hline
$x_m$ & Data symbol over $m$-th subcarrier\\\hline
$\varpi$ & Additive
white Gaussian noise with $\sigma^2$ variance\\\hline
$\gamma_m$ & Signal-to-noise-rato over $m$-th subcarrier\\\hline
\end{tabular}
\end{table}

The rest of the work can summarized as follows: Section two discusses the channel models of three links, i.e., from satellite to ground terminal, from satellite to RIS and from RIS to ground terminal. Section III represents problem formulation and proposed solution. Section IV provides simulation results and it discussion. Finally, Section V conclude this paper with potential future research topics. 

\textit{Notations}: The letters in both uppercase and lower case represents vectors and matrices. Besides that $\parallel$.$\parallel$ and $\mid$.$\mid$ show the Euclidean norm and absolute value. Moreover, diag$(.)$ denotes the diagonal matrix and $(.)^T$ is the conjugate transpose, respectively. The important symbols used in the channel modeling and problem formulation are depicted in Table I.

\section{System and Channel Models}
We consider a SatCom network, where a GEO satellite terminal intends to transmit the signal to the ground mobile terminal, as shown in Fig. \ref{Sat}. Due to the mobile ground terminal, the line-of-site connectivity from the satellite terminal is not always available. Thus, a RIS unit is also deployed on the surrounding building wall to assist the delivery of the satellite signal to the mobile ground terminal. We denote the satellite terminal by S, ground mobile terminal by G, and RIS unit by R, respectively. The R is equipped with $K$ reflecting elements and S follows multibeam technology. We consider that G receives signal from S through $M$ subcarriers using single antenna scenario \cite{9023370}. It is assumed that the channel state information is perfectly known in the system \cite{9543581}. In the following, we model realistic channels of different links such as S$\rightarrow$R, R$\rightarrow$D, and S$\rightarrow$D, respectively. The channels between S and R can be denoted by ${\textit{\textbf{g}}_{m,k}}\in\mathbb C^{K\times 1}$, such that ${\textit{\textbf{g}}_{m,k}}=\big[g_{m,1},g_{m,2},\dots,g_{m,k},\dots,g_{m,K}\big]$, where $g_{m,k}$ is the channel gain for the $m$-th subcarrier between S and R which can be expressed as
\begin{align}
g_{m,k}= \frac{\sqrt{G^S_{m}}}{4\pi \frac{d_2}{\lambda}}e^{-i\varphi},
\end{align}
where $G^S_{m}$ is the beam gain of S over $m$-th subcarrier, $d_2$ represents the distance from S to R over $m$-th subcarrier, and $\lambda$ denotes the wavelength. Here it is important to note that the beam gain does not change based on frequency carrier. What changes is the channel condition due to path loss. Moreover, $\varphi$ denotes the antenna phase of S. Accordingly, the channels between R and D over $m$-th subcarrier can be given by ${\textit{\textbf{h}}_{m,k}}\in\mathbb C^{K\times 1}$, such as ${\textit{\textbf{h}}_{m,k}}=\big[h_{m,1},h_{m,2},\dots,h_{m,k},\dots,h_{m,K}\big]$, where $h(m,k)$ denotes a channel from the $k$-th element of R towards D over $m$-th subcarrier and can be stated as \cite{8936989}
\begin{align}
h_{m,k}= \frac{\sqrt{G^D_{m}}X_RY_R}{d_3}\cos^2(\psi),
\end{align}
where $X_R$ and $Y_R$ denote the width and the length of the R, $G^D_m$ is the received antenna gain of D. $\psi\in[0,\pi/2]$ is the angle of incidence at the R while $d_3$ is the distance between R and D. Since we also consider the direct link from S to D, thus, the channels between S and D can be given as $f_m$, which can be defined as
\begin{align}
f_m= \frac{\sqrt{G^S_mG^D_m}}{4\pi\frac{d_1}{\lambda}},
\end{align}
where $d_1$ is the distance of direct link from S to D.
The received signal at D is the superposition of the direct signal from S to D and the reflected signal of R. The signal $x_m$ that D receives over $m$-th subcarrier can be expressed as
\begin{align}
y_m= \big({\textit{\textbf{g}}_{m,k}}{\bf \Theta}{\textit{\textbf{h}}_{m,k}}+f_m\big)^T\sqrt{p_{m}}x_m+\varpi_m,\label{y}
\end{align}
where ${\bf \Theta}\in\mathbb C^{K\times K}=\text{diag}(\alpha_1e^{j\theta_1},\alpha_2e^{j\theta_2},\dots, \alpha_Ke^{j\theta_K})$ is the matrix of reflection coefficients of R, where $\theta_k=[0,2\pi],$ denotes the phase shift and $\alpha_k=[0,1]$ is the amplitude of $k$-th reflector of R. Moreover, $p_m$ denotes the transmit power of S over the $m$-th subcarrier while $\varpi_m$ denotes the additive white Gaussian noise. Based on (\ref{y}), the received signal-to-noise-ratio (SNR) at D from S over $m$-th subcarrier can be expressed as
\begin{align}
\gamma_m=\frac{1}{\sigma^2}|({\textit{\textbf{g}}_{m,k}}{\bf \Theta}{\textit{\textbf{h}}_{m,k}}+f_m)^T p_m|^2,
\end{align} 
where $\sigma^2$ is the variance of additive white Gaussian noise.

\section{Proposed Channel Capacity Maximization Problem and Solution}
This section first formulates the optimization problem and then provides the detail of the proposed solution. In the considered framework, we aim to jointly optimize the transmit power of S over each subcarrier and reflect the phase shift of R to maximize the channel capacity of the system. The problem of channel capacity maximization can be formulated as:
\begin{alignat}{2}
\underset{{{\textit{\textbf{p}}},{\boldsymbol \theta}}}{\text{max}}&\ \ \sum\limits_{m=1}^M ||({\textit{\textbf{g}}_{m,k}}{\bf \Theta}{\textit{\textbf{h}}_{m,k}}+f_m)p_m||^2\label{6}\\
s.t. & \quad  |({\textit{\textbf{g}}_{m,k}}{\bf \Theta}{\textit{\textbf{h}}_{m,k}}+f_m)p_m|^2\geq \gamma^{min}_{m}\sigma^2,\ \forall m\in M, \tag{6a}\label{6a}\\
& \quad  p_m\geq 0,\ \forall m\in M,\ \sum\limits_{m=1}^Mp_m\leq p_t, \tag{6b}\label{6b}\\
& \quad  |\theta_k|\leq1,\ \forall k\in K, \tag{6c}\label{6c}
\end{alignat}
where $\textbf{p}=[p_1,p_2,\dots,p_M]$ and ${\boldsymbol \theta}=[\theta_1,\theta_2,\dots,\theta_K]$. Constraint (\ref{6a}) ensures the minimum transmission rate over each subcarrier, where $\gamma^{min}_{m}$ is the minimum SNR over $m$-th subcarrier. Constraint (\ref{6b}) controls the transmit power of S while (\ref{6c}) is the reflection phase shift constraint of R, respectively.

We can see that problem (\ref{6}) is coupled on two optimization variables ${\textit{\textbf{p}}},{\boldsymbol \theta}$, and nonconvex quadratically constrained quadratic programming problem, the traditional convex optimization method cannot be applied to solve this hard problem. Thus, we propose a new $\epsilon-$optimal algorithm which is based on MADS \cite{lakhmiri2021hypernomad}, which is an efficient method to solve the optimization problem (\ref{6}). This algorithm works based on exploration and exploitation to maximize the objective of (\ref{6}) iteratively. This work denotes it as a joint power allocation and reflecting phase shift design (JPAPS) algorithm, as depicted in algorithm 1. The proposed JPAPS algorithm has the ability to handle inequality constraints of various natures. It uses trial points which are evaluated through the function of a black box. Then the results obtained from these tests are further analyzed and used to generate new trial points. The iterative process of the JPAPS algorithm includes three phases, i) review phase, 2) search phase, and 3) poll phase. The proposed JPAPS algorithm performs the theoretical search using the search phrase based on one trial point, generated together with the previous successful route. 

\begin{algorithm}
	\SetAlgoLined
	\textbf{Step1:}\ Initialize $t \leftarrow 0,\varOmega^w_t > \varOmega^s_t > 0,z_t^s,z_t^w$\;
	$0 < {\varPhi^w} < {\varPhi^s} < 1$\;
	$F \leftarrow $ Solved \eqref{6}\;
	\textbf{Step2:}\
	\While{\text{Criterion not satisfied}}{
		$W_t = \underset{j\in \varPsi_t} \bigcup \{j + \varOmega_t^wC_t^w\}$\;
		$z_t^w = \left\{
		\begin{array}{l}
		\underset{j \in W_t}{\text{argmin}} \:\:  F \\
		\text{Subject to:} \:\: (6a)-(6c) \text{ of } \eqref{6} \\
		\end{array} \right.$\;
		\eIf{$F(z_t^w) > F(z_{t - 1}^w)$}{
			$S_t = \underset{j\in \varPsi_t} \bigcup \{j + \varOmega_t^sC_t^s\}$\;				
			$z_t^s = \left\{
			\begin{array}{l}
			\underset{j \in S_t}{\text{argmin}} \:\:  F \\
			\text{Subject to:} \:\: (6a)-(6c) \text{ of } \eqref{6} \\
			\end{array} \right.$\;
			\If{$F(z_t^s) > F(z_{t- 1}^s)$}{
				$\varOmega_{t+1}^s = \varOmega_t^s \cdot \varPhi^s$ \;	
			}
		}{
			$\varOmega_{t+1}^w = \frac{\varOmega_t^w}{\varPhi ^w}$\;
		}
		$t \leftarrow t+1$;
	}
	\caption{Joint power allocation and reflecting phase shift design (JPAPS) algorithm}\label{algo:JPTFA}
\end{algorithm}
The proposed JPAPS algorithm finds the optimal solution in the whole search space by applying the iterative contraction and expansion method. More specifically, this algorithm depends on the polls to decide the optimal solution for the current location of the iteration. Let $t$ denote the iteration index, which starts from zero. Moreover, $\varOmega^s_t$ shows the size of the poll at $t$-th iteration, and $\varOmega^w_t$ represents the width of the mesh at $t$-th iteration respectively. The mesh points at $t$-th iteration can be sated as $W_t$, which can be identified by placing the stencil at the current position $j$ and moving $\varOmega^w_t$ in compliance with the $C^w_t$. The set of $W$ consists of all previous and new points on the mesh. In particular, all previously visited points lying on the mesh and the new trial points around
any of the previously visited points using the instructions in $C^w_t$ at distances $\varOmega^w_t$. Using the set $W$, the optimization problem (\ref{6}) can be solved, and the results can be stored on set $z^w_{t}$. According to this process, the values of the objective function in problem (\ref{6}) (denoted as $F$) in the previous iterations $F(z^w_{t-1})$ should be compared with its updated values at current iterations $F(z^w_{t})$. If the values obtained by the previous iterations are smaller than the values of the updated iterations, it shows an improvement. However, if there is no improvement, the size of mesh will be increased by $\varPhi^w$ in the next iterations, i.e., $F(z^w_{t+1})$. The above process is also called an expansion. 

After improving the mesh analysis, the polling stage is initiated, where the polling steps can be defined by putting the stencil at location $j$ and moving $\varOmega^s_t$ according to the vector of direction mesh $C^s_t$. In mesh analysis, set $S$ is used to solve the optimization problem (\ref{6}), and the results are stored in set $z^s_t$, where $S$ is the set of polling points. According to this process, the values of the objective function in problem (\ref{6}) (denoted as $F$) in the previous iterations $F(z^s_{t+1})$ should be compared with its updated values at current iterations $F(z^s_{t})$. If the values are improvement, the poll size should be further reduced, i.e., $F(z^s_{t-1})$, by factor of $\varPhi^s$. This process is also called contraction. Algorithm 1 follows contraction as the objective values are improved and expands the search space when there is no further improvement record on the current mesh points. The said process will continue until the $\epsilon-$optimality is achieved.

\section{Numerical Results and Discussion}
In this section, we present and discuss the numerical results to show the performance of the proposed RIS-assisted SatCom network. We also check the benefits of RIS in SatCom by plotting a results of the proposed RIS-assisted SatCom versus the the conventional SatCom without RIS assistance. Unless stated, otherwise the simulation parameters are provided in Table II. Moreover, the results are obtained based on Monte Carlo simulations where $10^3$ realizations take the average of channels.

\begin{figure}[!t]
\centering
\includegraphics [width=0.52\textwidth]{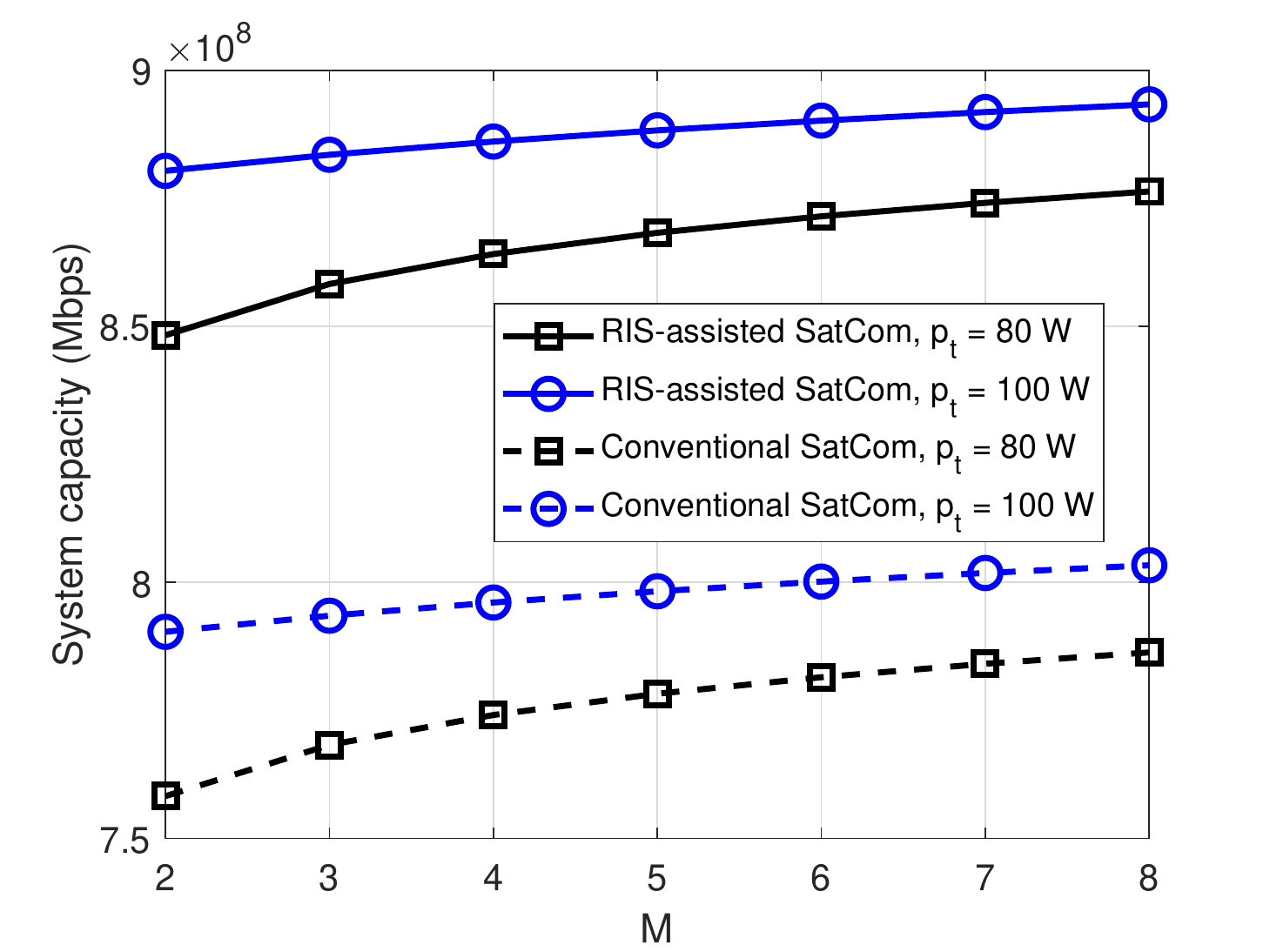}
\caption{Channel capacity of the system against the number of subcarriers with different transmit power of S, where the number of passive elements is $K=8$.}
\label{Fig1}
\end{figure}
\begin{table}
\centering
\caption{Simulation parameters}
\begin{tabular}{|c||c|} 
\hline 
Parameter & Definition  \\
\hline
Orbit & GEO\\\hline 
Carrier frequency & 1.5 GHz\\\hline 
Number of subcarriers $(M)$ & 8 \\\hline 
Bandwidth of each subcarrier $(B)$ & 10 MHz \\\hline 
Maximum power of S $(p_t)$ & 140 W \\\hline 
Noise variance $(\sigma^2)$ & 0.01  \\\hline 
Beam gain of S $(G^S_{m})$ & 51.8 dBi \\\hline 
Antenna gain of D $(G^D_m)$ & -13.5 dB/K\\\hline 
Angle of incidence at R $(\psi)$ & $60^{\circ}$ \\\hline 
Height of R (m) & 100\\\hline 
Maximum elements at R $(K)$ & 10\\\hline 
Width and Length of R, i.e., $X_R$, $Y_R$ (m) & 1\\\hline
Reflection coefficient of R & 1\\
\hline 
\end{tabular}
\end{table}
\begin{figure}[!t]
\centering
\includegraphics [width=0.52\textwidth]{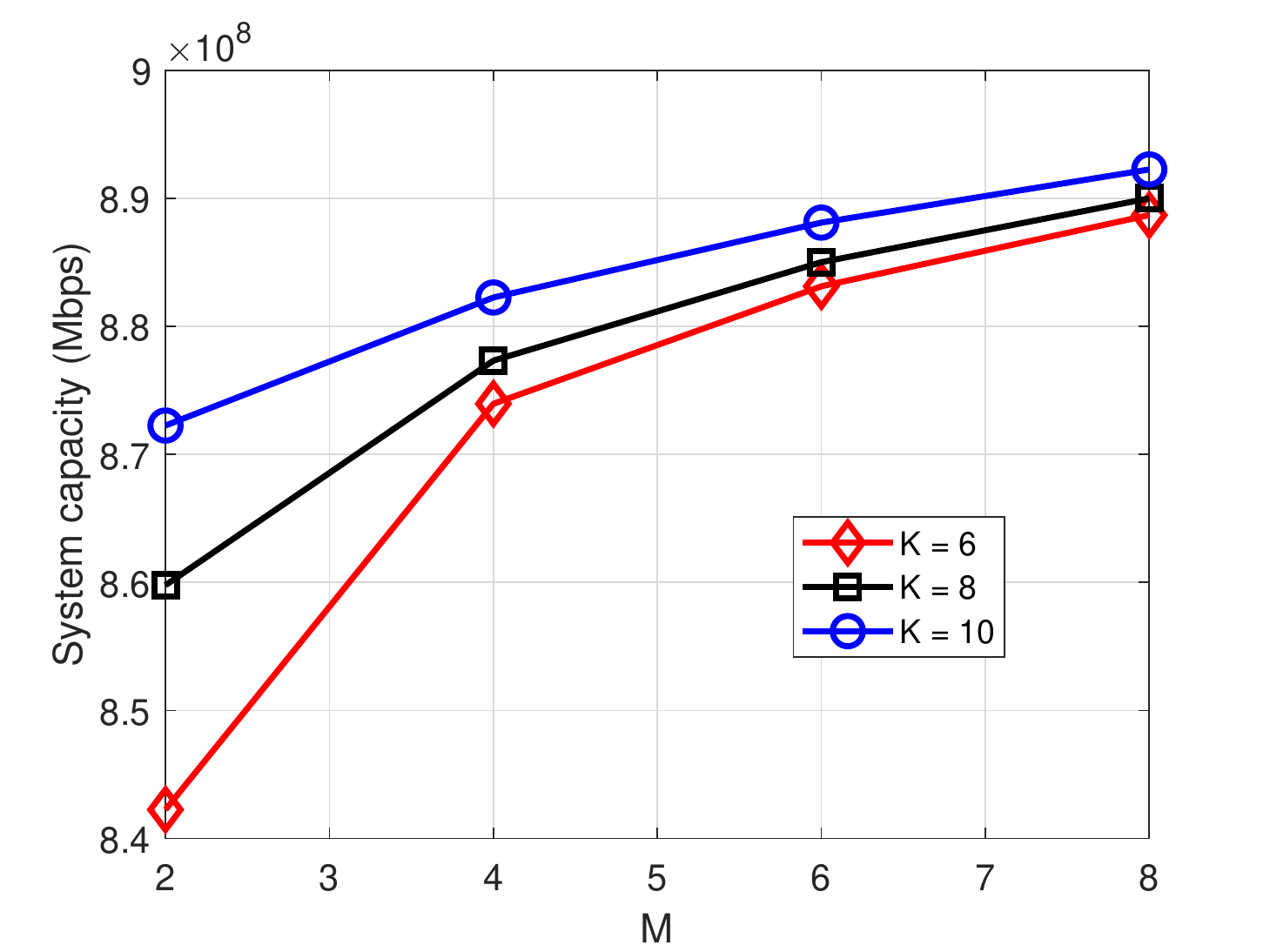}
\caption{Channel capacity of the system against the number of subcarriers with different number of passive elements at RIS, where the transmit power of S is set as $p_t=100$ W.}
\label{Fig2}
\end{figure}
\subsection{RIS-Assisted SatCom Versus Conventional SatCom}
Fig. \ref{Fig1} compares the proposed RIS-assisted SatCom with conventional SatCom without RIS assistance. This figure plots the system channel capacity against the number of subcarriers where the transmit power of S is set as $p_t=80$ W and $p_t=100$ W, respectively. It can be seen that the channel capacity of the system for both RIS-assisted and conventional SatCom increases as the number of subcarriers varies from $M=1$ to $M=8$ is increase. Besides that the systems with more transmit power achieve a high channel capacity compared to the systems with lower transmit power at S. We can evidence that the proposed RIS-assisted SatCom significantly outperforms the conventional SatCom without RIS assistance. For instance, with equal system parameters, when the number of subcarriers is $M=8$ and the transmit power of S is $p_t=100$, the channel capacity of the proposed RIS-assisted SatCom is near to $8.9\times10^8$ Mbps while the channel capacity of the conventional SatCom without RIS assistance is around $8\times10^8$ Mbps only. It indicates the important role of RIS in next-generation high throughput SatCom networks. 

\subsection{Impact of System Parameters on RIS-Assisted SatCom}
Fig. \ref{Fig2} shows the performance of the proposed RIS-assisted SatCom framework against the number of subcarriers for different number of passive reflecting elements. In this figure, the subcarriers vary from $M=2$ to $M=8$, and the number of passive reflecting elements is set as $K=6$, $K=8$, and $K=10$, respectively. We can observe that the channel capacity of the proposed optimization framework improves as the number of subcarriers and passive reflecting elements increase. One can also be noted that the performance gap among different curves gradually narrows down as the number of subcarriers increases. It indicates that the benefits are most substantial in the system with few subcarriers and more reflecting elements. In general, this is a scenario that is practically appealing because the installment of the more reflecting elements is cheaper than using extra radio resources. 
\begin{figure}[!t]
\centering
\includegraphics [width=0.52\textwidth]{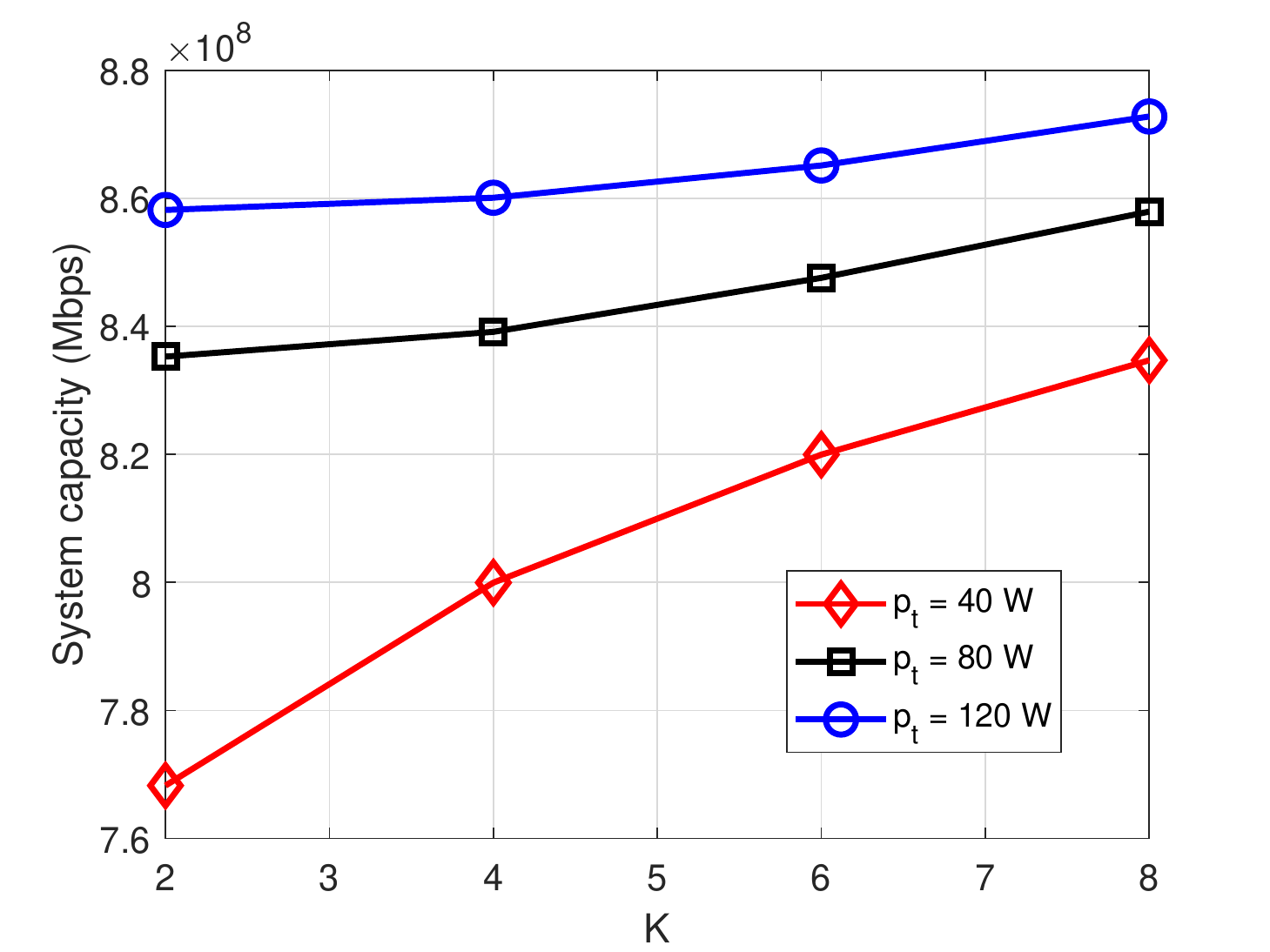}
\caption{Channel capacity of the system against the number of passive reflecting elements at R with different transmit power of S, where the number of subcarriers is set as $M=8$.}
\label{Fig3}
\end{figure}

Fig. \ref{Fig3} illustrates the impact of the number of passive reflecting elements on the system channel capacity of SatCom network. Here the number of reflecting elements at R varies from 2 to 8 and the transmit power at S is set as 40, 80, and 120 W, respectively. As depicted in the figure, the system channel capacity with more available transmit power is higher than that of the other systems with comparatively lower transmit power. It can also be seen that the system with more passive reflecting elements offers a high channel capacity compared to the other systems with a lower number of passive reflecting elements. It is because more reflecting elements at R can enhance the effective gain of the reflecting path.  

\begin{figure}[!t]
\centering
\includegraphics [width=0.52\textwidth]{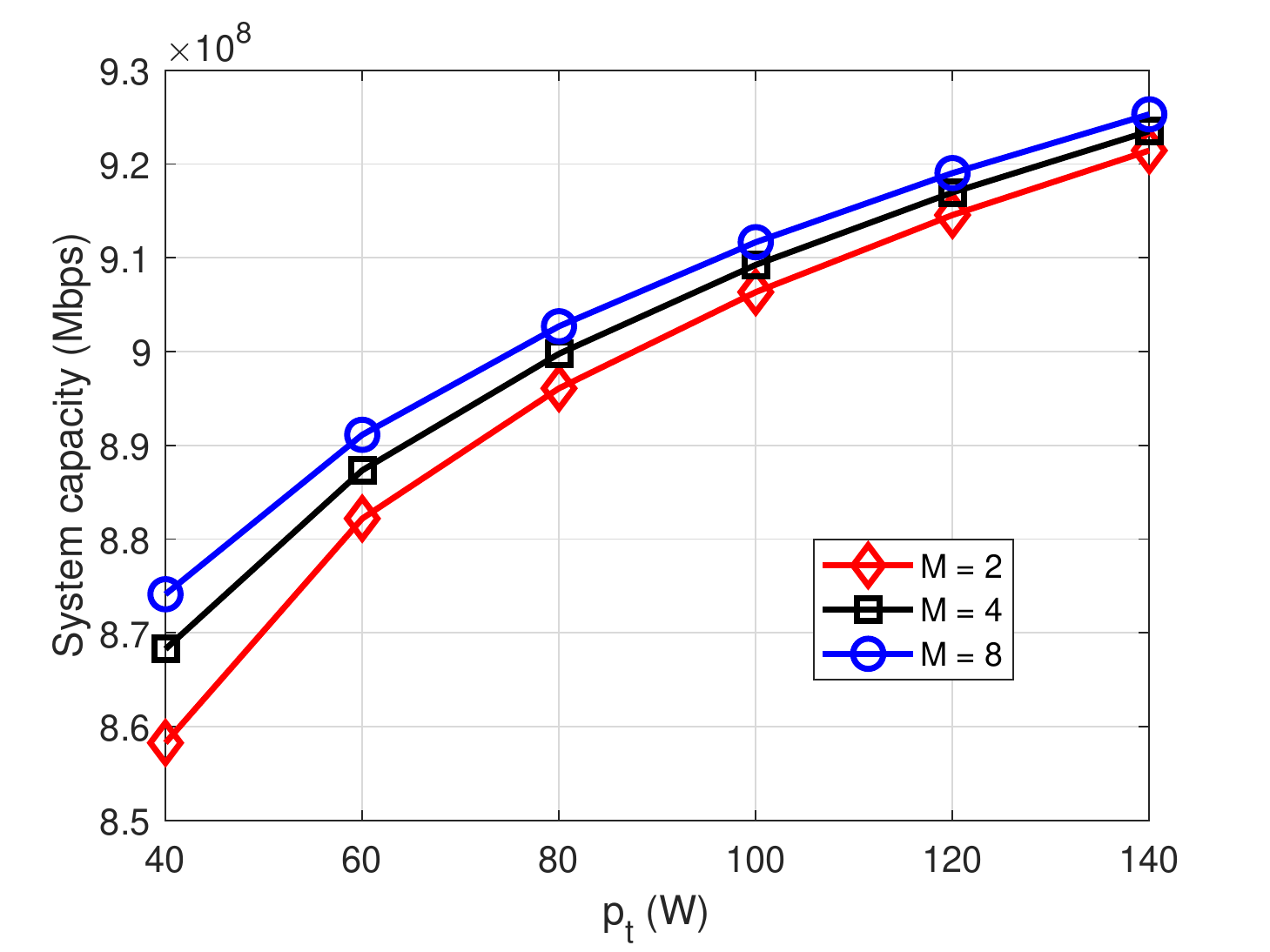}
\caption{Channel capacity of the system against the increasing transmit power of S with different number of subcarriers, where the number of passive reflecting elements at R is set as $K=16$.}
\label{Fig4}
\end{figure}

Finally, Fig. \ref{Fig4} depicts the system channel capacity versus the increasing power of S for the different number of subcarriers. In this figure, the transmit power varies from 40 W to 120 W and the number of subcarriers is set as 2, 4, and 8. Once again, we can see that the performance of the proposed system improves as the available transmit power increases. This is because the channel capacity is a variant of Shannon capacity. It can also be evident that the system with more subcarriers achieves a high transmission rate compared to those with comparatively lower subcarriers. This is because the system with more subcarriers can transmit more information which improves the overall channel capacity.

\section{Conclusion}
Being a very low cast architecture, RIS has recently attracted significant attention in both academia and industry due to the high spectral and energy efficiency. On the other side, the demand of high data rate and massive connectivity is increasing rapidly in next-generation SatCom networks. This paper has provided a joint optimization framework for RIS-assisted SatCom networks. Specifically, the signal delivery has also been assisted by RIS along with the direct link from satellite to the ground mobile terminal. The transmit power of satellite over multiple subcarrier and phase shift design over different reflecting element at RIS have been optimized. The objective has to maximized the channel capacity of the system while taken into account the quality of services over each subcarrier. The joint optimization has been formulated as nonconvex quadratically constrained quadratic programing problem and hard to solve. The efficient solution has been obtained through MADS algorithm referred as JPAPS. Numerical results show that the proposed RIS-assisted SatCom optimization framework has the advantage over the conventional SatCom framework without RIS assistance. Our work can be easily extended from one ground terminal to multiple ground terminals. In that case, a subcarrier cannot be associated to more than 1 ground terminal and a ground terminal can have access to multiple subcarriers at any given time. Moreover, this framework can be easily applied to other non-GEO SatCom networks. These interesting yet explored topics will be performed in the future.
\section{Acknowledgement}
This work was supported by Luxembourg National Research Fund (FNR) under the CORE project RISOTTI C20/IS/14773976.

\bibliographystyle{IEEEtran}
\bibliography{Wali_EE}

\end{document}